\documentclass[twocolumn,aps,prx,amsmath,amssymb,
 superscriptaddress,nofootinbib,reprint,]{revtex4-1}
\usepackage[T1]{fontenc}
\usepackage{stmaryrd}
\usepackage{color}
\usepackage{array}
\usepackage{enumitem}
\usepackage{mathpazo}
\usepackage{setspace}
\usepackage{bm}
\usepackage{amssymb}
\usepackage{amsthm}
\usepackage{mathtools}
\usepackage{multirow}
\usepackage{cases}
\usepackage[colorlinks=true,urlcolor=blue,citecolor=blue,linkcolor=blue]{hyperref}
\usepackage{pgfplots}
\usepackage{lipsum}
\usepackage{youngtab}
\usepackage{subfigure}
\usepackage{algorithm}
\usepackage{algpseudocode}

\usepackage[capitalise]{cleveref}
\usepackage{booktabs}

\newtheorem{definition}{Definition}
\newtheorem{proposition}[definition]{Proposition}
\newtheorem{lemma}[definition]{Lemma}

\newtheorem{theorem}[definition]{Theorem}

\newtheorem{corollary}[definition]{Corollary}
\newtheorem{conjecture}[definition]{Conjecture}

\newtheorem{remark}[definition]{Remark}
\newtheorem{example}[definition]{Example}
\newtheorem{question}[definition]{Question}

\def\Dbar{\leavevmode\lower.6ex\hbox to 0pt
{\hskip-.23ex\accent"16\hss}D}
\makeatletter
\def\url@leostyle{%
  \@ifundefined{selectfont}{\def\UrlFont{\sf}}{\def\UrlFont{\small\ttfamily}}}
\makeatother
\DeclareMathOperator{\tr}{tr} %

\def\bcj{\begin{conjecture}}
\def\ecj{\end{conjecture}}
\def\bcr{\begin{corollary}}
\def\ecr{\end{corollary}}
\def\bd{\begin{definition}}
\def\ed{\end{definition}}
\def\bea{\begin{eqnarray}}
\def\eea{\end{eqnarray}}
\def\bem{\begin{enumerate}}
\def\eem{\end{enumerate}}
\def\bex{\begin{example}}
\def\eex{\end{example}}
\def\bim{\begin{itemize}}
\def\eim{\end{itemize}}
\def\bl{\begin{lemma}}
\def\el{\end{lemma}}
\def\bpf{\begin{proof}}
\def\epf{\end{proof}}
\def\bpp{\begin{proposition}}
\def\epp{\end{proposition}}
\def\bqu{\begin{question}}
\def\equ{\end{question}}
\def\br{\begin{remark}}
\def\er{\end{remark}}
\def\bt{\begin{theorem}}
\def\et{\end{theorem}}

\def\btb{\begin{tabular}}
\def\etb{\end{tabular}}

\newcommand{\nc}{\newcommand}

\def\i{\iota}

 \nc{\bA}{{\bf A}} \nc{\bB}{{\bf B}} \nc{\bC}{{\bf C}}
 \nc{\bD}{{\bf D}} \nc{\bE}{{\bf E}} \nc{\bF}{{\bf F}}
 \nc{\bG}{{\bf G}} \nc{\bH}{{\bf H}} \nc{\bI}{{\bf I}}
 \nc{\bJ}{{\bf J}} \nc{\bK}{{\bf K}} \nc{\bL}{{\bf L}}
 \nc{\bM}{{\bf M}} \nc{\bN}{{\bf N}} \nc{\bO}{{\bf O}}
 \nc{\bP}{{\bf P}} \nc{\bQ}{{\bf Q}} \nc{\bR}{{\bf R}}
 \nc{\bS}{{\bf S}} \nc{\bT}{{\bf T}} \nc{\bU}{{\bf U}}
 \nc{\bV}{{\bf V}} \nc{\bW}{{\bf W}} \nc{\bX}{{\bf X}}
 \nc{\bZ}{{\bf Z}}

\nc{\cA}{{\cal A}} \nc{\cB}{{\cal B}} \nc{\cC}{{\cal C}}
\nc{\cD}{{\cal D}} \nc{\cE}{{\cal E}} \nc{\cF}{{\cal F}}
\nc{\cG}{{\cal G}} \nc{\cH}{{\cal H}} \nc{\cI}{{\cal I}}
\nc{\cJ}{{\cal J}} \nc{\cK}{{\cal K}} \nc{\cL}{{\cal L}}
\nc{\cM}{{\cal M}} \nc{\cN}{{\cal N}} \nc{\cO}{{\cal O}}
\nc{\cP}{{\cal P}} \nc{\cQ}{{\cal Q}} \nc{\cR}{{\cal R}}
\nc{\cS}{{\cal S}} \nc{\cT}{{\cal T}} \nc{\cU}{{\cal U}}
\nc{\cV}{{\cal V}} \nc{\cW}{{\cal W}} \nc{\cX}{{\cal X}}
\nc{\cZ}{{\cal Z}}

\nc{\hA}{{\hat{A}}} \nc{\hB}{{\hat{B}}} \nc{\hC}{{\hat{C}}}
\nc{\hD}{{\hat{D}}} \nc{\hE}{{\hat{E}}} \nc{\hF}{{\hat{F}}}
\nc{\hG}{{\hat{G}}} \nc{\hH}{{\hat{H}}} \nc{\hI}{{\hat{I}}}
\nc{\hJ}{{\hat{J}}} \nc{\hK}{{\hat{K}}} \nc{\hL}{{\hat{L}}}
\nc{\hM}{{\hat{M}}} \nc{\hN}{{\hat{N}}} \nc{\hO}{{\hat{O}}}
\nc{\hP}{{\hat{P}}} \nc{\hR}{{\hat{R}}} \nc{\hS}{{\hat{S}}}
\nc{\hT}{{\hat{T}}} \nc{\hU}{{\hat{U}}} \nc{\hV}{{\hat{V}}}
\nc{\hW}{{\hat{W}}} \nc{\hX}{{\hat{X}}} \nc{\hZ}{{\hat{Z}}}

\newcommand{\bra}[1]{\langle#1|}
\newcommand{\ket}[1]{|#1\rangle}

\def\Dbar{\leavevmode\lower.6ex\hbox to 0pt
{\hskip-.23ex\accent"16\hss}D}

\usepackage{balance}

\begin{document}

\def\be{\begin{eqnarray}}
\def\ee{\end{eqnarray}}


\newcommand{\ca}{\mathcal A}

\newcommand{\cb}{\mathcal B}
\newcommand{\cc}{\mathcal C}
\newcommand{\cd}{\mathcal D}
\newcommand{\ce}{\mathcal E}
\newcommand{\cf}{\mathcal F}
\newcommand{\cg}{\mathcal G}
\newcommand{\ch}{\mathcal H}
\newcommand{\ci}{\mathcal I}
\newcommand{\cj}{\mathcal J}
\newcommand{\ck}{\mathcal K}
\newcommand{\cl}{\mathcal L}
\newcommand{\cm}{\mathcal M}
\newcommand{\cn}{\mathcal N}
\newcommand{\co}{\mathcal O}
\newcommand{\cp}{\mathcal P}
\newcommand{\cq}{\mathcal Q}
\newcommand{\calr}{\mathcal R}
\newcommand{\cs}{\mathcal S}
\newcommand{\ct}{\mathcal T}
\newcommand{\cu}{\mathcal U}
\newcommand{\cv}{\mathcal V}
\newcommand{\cw}{\mathcal W}
\newcommand{\cx}{\mathcal X}
\newcommand{\cy}{\mathcal Y}
\newcommand{\cz}{\mathcal Z}


\newcommand{\sa}{\mathscr{A}}
\newcommand{\sm}{\mathscr{M}}


\newcommand{\fa}{\mathfrak{a}}  \newcommand{\Fa}{\mathfrak{A}}
\newcommand{\fb}{\mathfrak{b}}  \newcommand{\Fb}{\mathfrak{B}}
\newcommand{\fc}{\mathfrak{c}}  \newcommand{\Fc}{\mathfrak{C}}
\newcommand{\fd}{\mathfrak{d}}  \newcommand{\Fd}{\mathfrak{D}}
\newcommand{\fe}{\mathfrak{e}}  \newcommand{\Fe}{\mathfrak{E}}
\newcommand{\ff}{\mathfrak{f}}  \newcommand{\Ff}{\mathfrak{F}}
\newcommand{\fg}{\mathfrak{g}}  \newcommand{\Fg}{\mathfrak{G}}
\newcommand{\fh}{\mathfrak{h}}  \newcommand{\Fh}{\mathfrak{H}}
\newcommand{\fraki}{\mathfrak{i}}       \newcommand{\Fraki}{\mathfrak{I}}
\newcommand{\fj}{\mathfrak{j}}  \newcommand{\Fj}{\mathfrak{J}}
\newcommand{\fk}{\mathfrak{k}}  \newcommand{\Fk}{\mathfrak{K}}
\newcommand{\fl}{\mathfrak{l}}  \newcommand{\Fl}{\mathfrak{L}}
\newcommand{\fm}{\mathfrak{m}}  \newcommand{\Fm}{\mathfrak{M}}
\newcommand{\fn}{\mathfrak{n}}  \newcommand{\Fn}{\mathfrak{N}}
\newcommand{\fo}{\mathfrak{o}}  \newcommand{\Fo}{\mathfrak{O}}
\newcommand{\fp}{\mathfrak{p}}  \newcommand{\Fp}{\mathfrak{P}}
\newcommand{\fq}{\mathfrak{q}}  \newcommand{\Fq}{\mathfrak{Q}}
\newcommand{\fr}{\mathfrak{r}}  \newcommand{\Fr}{\mathfrak{R}}
\newcommand{\fs}{\mathfrak{s}}  \newcommand{\Fs}{\mathfrak{S}}
\newcommand{\ft}{\mathfrak{t}}  \newcommand{\Ft}{\mathfrak{T}}
\newcommand{\fu}{\mathfrak{u}}  \newcommand{\Fu}{\mathfrak{U}}
\newcommand{\fv}{\mathfrak{v}}  \newcommand{\Fv}{\mathfrak{V}}
\newcommand{\fw}{\mathfrak{w}}  \newcommand{\Fw}{\mathfrak{W}}
\newcommand{\fx}{\mathfrak{x}}  \newcommand{\Fx}{\mathfrak{X}}
\newcommand{\fy}{\mathfrak{y}}  \newcommand{\Fy}{\mathfrak{Y}}
\newcommand{\fz}{\mathfrak{z}}  \newcommand{\Fz}{\mathfrak{Z}}

\newcommand{\cfg}{\dot \fg}
\newcommand{\cFg}{\dot \Fg}
\newcommand{\ccg}{\dot \cg}
\newcommand{\circj}{\dot {\mathbf J}}
\newcommand{\circs}{\circledS}
\newcommand{\jmot}{\mathbf J^{-1}}


\newcommand{\rmd}{\mathrm d}
\newcommand{\mca}{\ ^-\!\!\ca}
\newcommand{\pca}{\ ^+\!\!\ca}
\newcommand{\peq}{^\Psi\!\!\!\!\!=}
\newcommand{\lt}{\left}
\newcommand{\rt}{\right}
\newcommand{\HN}{\hat{H}(N)}
\newcommand{\HM}{\hat{H}(M)}
\newcommand{\Hv}{\hat{H}_v}
\newcommand{\cyl}{\mathbf{Cyl}}
\newcommand{\lag}{\left\langle}
\newcommand{\rag}{\right\rangle}
\newcommand{\Ad}{\mathrm{Ad}}
\newcommand{\trace}{\mathrm{tr}}
\newcommand{\bbc}{\mathbb{C}}
\newcommand{\AC}{\overline{\mathcal{A}}^{\mathbb{C}}}
\newcommand{\Ar}{\mathbf{Ar}}
\newcommand{\uc}{\mathrm{U(1)}^3}
\newcommand{\M}{\hat{\mathbf{M}}}
\newcommand{\spin}{\text{Spin(4)}}
\newcommand{\id}{\mathrm{id}}
\newcommand{\Pol}{\mathrm{Pol}}
\newcommand{\Fun}{\mathrm{Fun}}
\newcommand{\bp}{p}
\newcommand{\act}{\rhd}
\newcommand{\data}{\lt(j_{ab},A,\bar{A},\xi_{ab},z_{ab}\rt)}
\newcommand{\datao}{\lt(j^{(0)}_{ab},A^{(0)},\bar{A}^{(0)},\xi_{ab}^{(0)},z_{ab}^{(0)}\rt)}
\newcommand{\deltadata}{\lt(j'_{ab}, A',\bar{A}',\xi_{ab}',z_{ab}'\rt)}
\newcommand{\background}{\lt(j_{ab}^{(0)},g_a^{(0)},\xi_{ab}^{(0)},z_{ab}^{(0)}\rt)}
\newcommand{\sgn}{\mathrm{sgn}}
\newcommand{\vth}{\vartheta}
\newcommand{\rmi}{\mathrm{i}}
\newcommand{\bfmu}{\pmb{\mu}}
\newcommand{\bfnu}{\pmb{\nu}}
\newcommand{\bfm}{\mathbf{m}}
\newcommand{\bfn}{\mathbf{n}}
\newcommand{\perk}{\mathfrak{S}_k}
\newcommand{\dens}{\mathrm{D}}
\newcommand{\iden}{\mathbb{I}}
\newcommand{\End}{\mathrm{End}}
\newcommand{\C}{\mathbb{C}}


\newcommand{\sz}{\mathscr{Z}}
\newcommand{\sk}{\mathscr{K}}

\title{A variational quantum algorithm for Hamiltonian diagonalization}

\author{Jinfeng Zeng}
\email[]{jfzeng@ust.hk}
\affiliation{Department of Physics, The Hong Kong University of Science and Technology, Clear Water Bay, Kowloon, Hong Kong}

\author{Chenfeng Cao}
\email[]{chenfeng.cao@connect.ust.hk}
\affiliation{Department of Physics, The Hong Kong University of Science and Technology, Clear Water Bay, Kowloon, Hong Kong}

\author{Chao Zhang}
\affiliation{Peng Cheng Laboratory, Shenzhen, 518055, China}

\author{Pengxiang Xu}
\affiliation{Peng Cheng Laboratory, Shenzhen, 518055, China}

\author{Bei Zeng}
\email[]{zengb@ust.hk}
\affiliation{Department of Physics, The Hong Kong University of Science and Technology, Clear Water Bay, Kowloon, Hong Kong}

\date{\today}

\begin{abstract}
Hamiltonian diagonalization is at the heart of understanding physical properties and practical applications of 
quantum systems. It is highly desired to design quantum algorithms that can speedup Hamiltonian diagonalization,
especially those can be implemented on near-term quantum devices. 
In this work, we propose a variational algorithm for Hamiltonians diagonalization (VQHD) of quantum systems, which explores the important physical properties, such as temperature, locality, and correlation, of the system. The key idea
is that the thermal states of the system encode the information of eigenvalues and eigenstates of the system Hamiltonian.
To obtain the full spectrum of the Hamiltonian, we use a quantum imaginary time evolution algorithm with high temperature, which prepares a thermal state with a small correlation length. With Trotterization, this then allows us to implement each step of imaginary time evolution by a local unitary transformation on only a small number of sites. 
Diagonalizing these thermal states hence leads to a full knowledge of the Hamiltonian eigensystem. We apply our algorithm to diagonalize local Hamiltonians and return results with high precision. Our VQHD algorithm sheds new light on the applications of near-term quantum computers.
\end{abstract}

\maketitle
\renewcommand\theequation{\arabic{section}.\arabic{equation}}
\setcounter{tocdepth}{4}
\makeatletter
\@addtoreset{equation}{section}
\makeatother

\section{Introduction}

Naturally arising Hamiltonian of quantum systems exhibit local structure,
which allows efficient algorithms on quantum computers to simulate the evolution
of these systems. Diagonalizing these Hamiltonians, however, is a more challenging
task for quantum computing, which also serves as important subroutines, for instance, 
of the celebrated density functional theory (DFT)~\cite{Hohenberg1964DFT, Kohn1965DFT}  for important applications of quantum simulation
in chemistry, materials sciences and technologies~\cite{baker2020density}. Quantum algorithms have been developed for
finding the eigenvalues and eigenstates of Hamiltonians,  
for instance the one based on quantum fast Fourier transform~\cite{abrams1999quantum}.
However, implementing these algorithms require fault-tolerance~\cite{gaitan2008quantum}, which 
are not expected to be achievable in the near future.

With the current noisy intermedia-scale quantum (NISQ) era~\cite{Preskill2018}, it is highly desired to 
design quantum algorithms that can take advantage of the near-term
quantum devices. Many variational/hybrid quantum algorithms are proposed in recent years,
for tasks such as finding ground states of Hamiltonians (variational quantum eigen solver, 
VQE~\cite{peruzzo2014variational,McClean2016VQE,Kandala2017}),
finding approximate solutions to combinatorial optimization problems (quantum approximate optimization algorithm QAOA~\cite{farhi2014quantum});
finding singular values of matrices~\cite{wang2020Variational}, pure state Schmidt decomposition~\cite{bravo2020quantum},
and training quantum Boltzmann machine (variational quantum Boltzmann machine~\cite{zoufal2020variational,shingu2020boltzmann}).
Along this line, the variational algorithm for finding Hamiltonian spectra has also been discussed~\cite{jones2019variational}, with also various new methods for finding excited states~\cite{higgott2019variational,nakanishi2019subspace}.

Hamiltonian diagonalization is to find all the excited states and ground states. Ref.~\cite{higgott2019variational} proposes a variational quantum deflation (VQD) algorithm to calculate the excited states, which extends the original VQE by imposing orthogonality conditions between the ansatz states. The orthogonality condition of the $k$-th excited state needs to estimate the inner products between the $k$-th ansatz state and all of the lower $(k-1)$  ground/excited states. The estimation of the overlap in VQD is not easily implementable on the NISQ devices. Ref.~\cite{nakanishi2019subspace} propose a subspace search variational quantum eigensolver (SSVQE), which get rid of many overlap estimations but increase the circuit depth.  Additionally, there are variational methods~\cite{cerezo2020variational, larose2019variational} for diagonalizing quantum state instead of the hamiltonian. The variational quantum state diagonalization (VQSD)~\cite{larose2019variational} method also need to calculate overlap and use the diagonalized inner product test to estimate the overlap instead of doubling the ansatz circuit depth in VQD. The VQSD saves the circuit depth but increases the number of qubits. Further, another method for diagonalizing density matrix called variational quantum state eigensolver (VQSE)~\cite{cerezo2020variational} do not need to estimate the overlap, which requires only $n$ qubits as the same with the original quantum state. 


In this work, we propose a new variational quantum algorithm for Hamiltonian diagonalization (VQHD) for quantum systems.
The key idea is that for any system Hamiltonian $H$, the thermal state $\rho_{\beta}=e^{-\beta H}/\tr (e^{-\beta H})$
encodes the information of the eigenvalues and eigenvectors of $H$, where $\beta=1/k_{B}T$ with $T$ the temperature of 
the system. For small $\beta$, $\rho_{\beta}$ is full rank and diagonalizing $\rho_{\beta}$ directly returns the eigensystem of $H$. Hence if we prepare the thermal state $\rho_{\beta}$, we can then use variational algorithms to diagonalize $\rho_{\beta}$, for obtaining the eigensystem of $H$.  

To prepare the thermal state $\rho_{\beta}$, one can apply an imaginary time evolution with a thermofield double state, as proposed in~\cite{wu2019variational}. The idea is illustrated in Fig.~\ref{qite-locality} (a), where each connected pair of dots represents a two-qubit maximally entangled 
state, and the initial thermofield double state $\ket{\text{TFD}(0)}$ is hence a product of $n$ entangled pairs. The imaginary time evolution $e^{-\beta H/2}$ on $\ket{\text{TFD}(0)}$ returns the state $\ket{\text{TFD}(\beta)}$, and $\rho_{\beta}$ will be then obtained on the bottom $n$ qubits by tracing out the top $n$ qubits. For applying $e^{-\beta H/2}$, we choose a quantum imaginary time evolution (QITE) algorithm as proposed in~\cite{motta2020determining}. An advantage of this choice is that, for small $\beta$, the many-body state $\rho_{\beta}$ exhibits small correlation length, local unitary transformations on a relatively small size of local sites hence suffices to simulate the imaginary time evolution on a quantum computer. We then use VQSE to diagonalize the $\rho_{\beta}$, which do not need to estimate the overlaps. Comparing to directly diagonalize the Hamiltonian using variational quantum algorithms~\cite{cerezo2020variational, larose2019variational}, our method makes use of the short depth circuit in the preparation of the thermal state. Moreover, we propose an alternative variational quantum algorithm for hamiltonian diagonalization on NISQ devices.

We apply our algorithm for diagonalizing various local Hamiltonians, to obtain the full spectrum. Notice that a larger value of $\beta$ can suppress high energy states which will then return low-lying eigenstates of $H$. Depending on the situation, our algorithm hence can also be applied to find low-lying states. Our method adds a new tool to the family of NISQ algorithms and will shed light on the near-term application of quantum computers.


\section{The variational algorithm for Hamiltonian diagonalization}

In this section, we present our VQHD algorithm and simulation results.

\subsection{The VQHD algoritm}

Consider a quantum system of $n$ qubits. The system Hamiltonian $H$ adopts the form
\begin{equation}
H=\sum_i h[i],
\end{equation}
with each $h[i]$ acting nontrivially on geometrically local sites.  

Instead of directly diagonalizing the Hamiltonian, we diagonalize the thermal state
\begin{eqnarray}
H |\psi_k \rangle = \lambda_k | \psi_k \rangle \rightarrow  \rho_{\beta} | \psi_k \rangle =  \lambda'_k | \psi_k \rangle
\end{eqnarray}
The thermal state $\rho_{\beta}$ of the system has the form
\begin{equation}
\rho_{\beta}=\frac{e^{-\beta H}}{\tr(e^{-\beta H})},
\end{equation}
where $\beta=\frac{1}{k_{B} T}$, and $T$ is the temperature of the system. The thermal state $\rho_{\beta}$ shares the same eigen states with the original Hamiltonian $H$.  And the eigen value of thermal state is $\lambda'_k = \frac{e^{ -\beta \lambda_k}}{\sum_k e^{-\beta \lambda_k} }$.

Our VQHD algorithm is summarized in Algorithm~\ref{alg:VQHD}. 
\begin{algorithm}[H]
	\begin{algorithmic}	
      \caption{Variational algorithm for Hamiltonian diagonalization.}
        \Require The Hamiltonian $H$ and the inverse temperature $\beta$.  The parameter $\Delta\tau$, $D$ for QITE to prepare the thermal state. A variational quantum circuit ansatz $V(\boldsymbol{\theta})$ for diagonal quantum state. The learning rate $\eta$.
        
        \Ensure  eigen states $\{| \psi_k \rangle\}$ and eigen values $\{\lambda_k \}$.
         \State {\color{blue}{\texttt{\# (S1) Prepare the  thermal state $\rho_{\beta}$.}}}
	      \State Perpare initial state $\ket{\phi_0}  = \frac{1}{\sqrt{2^n}} \sum_{i=1}^{2^n}\ket{i}\ket{i}$ with  $2n$ qubits.
         \State Implement $\ket{\text{TFD}(\beta)} =  \sqrt{\frac{2^n}{\tr(e^{-\beta H})}} e^{-\beta H/2} \ket{\phi_0}$ with QITE.
          \State Trace the last $n$ qubits $\rho_{\beta}=\tr_{n+1,\ldots ,2n}\ket{\text{TFD}(\beta)}\bra{\text{TFD}(\beta)}$.
          
          \State {\color{blue}{\texttt{\# (S2) Diagonalize the  thermal state $\rho_{\beta}$.}}}
        
	    \While{$\boldsymbol{\theta}$ have not converged}
         \State Apply the quantum circuit $\rho_{f} = V(\boldsymbol{\theta}) \rho_{\beta} V^{\dagger}(\boldsymbol{\theta}) $.
         \State Compute the loss function $C(\boldsymbol{\theta}) =\operatorname{Tr}\left[H_{cost}\rho_{f}\right]$.
        \State Compute the gradient of loss function $\nabla_{\boldsymbol{\theta}} C(\boldsymbol{\theta})$.
        \State $\boldsymbol{\theta}=\boldsymbol{\theta} -\eta \nabla_{\boldsymbol{\theta}} C(\boldsymbol{\theta})$.
         \EndWhile
         \State Get the optimal $\boldsymbol{\theta^{*}}$.
         \State Compute the eigen states $\{| \psi_k \rangle\ = V^{\dagger}(\boldsymbol{\theta^{*}} )| k \rangle \}$ .
          \State Compute the eigen values $\{\lambda_k =  \langle k| V(\boldsymbol{\theta^*})  H  V^{\dagger}(\boldsymbol{\theta^{*}} )| k \rangle \}$.
		\label{alg:VQHD}
	\end{algorithmic}
\end{algorithm}

\subsection{Algorithm details}

In {\bf Algorithm 1}, there are two major steps (S1) and (S2). There are various methods one can use for each (S1) and (S2). We discuss the details in the following.


\begin{figure*}[tb]
\centerline{\includegraphics[width=0.95\textwidth]{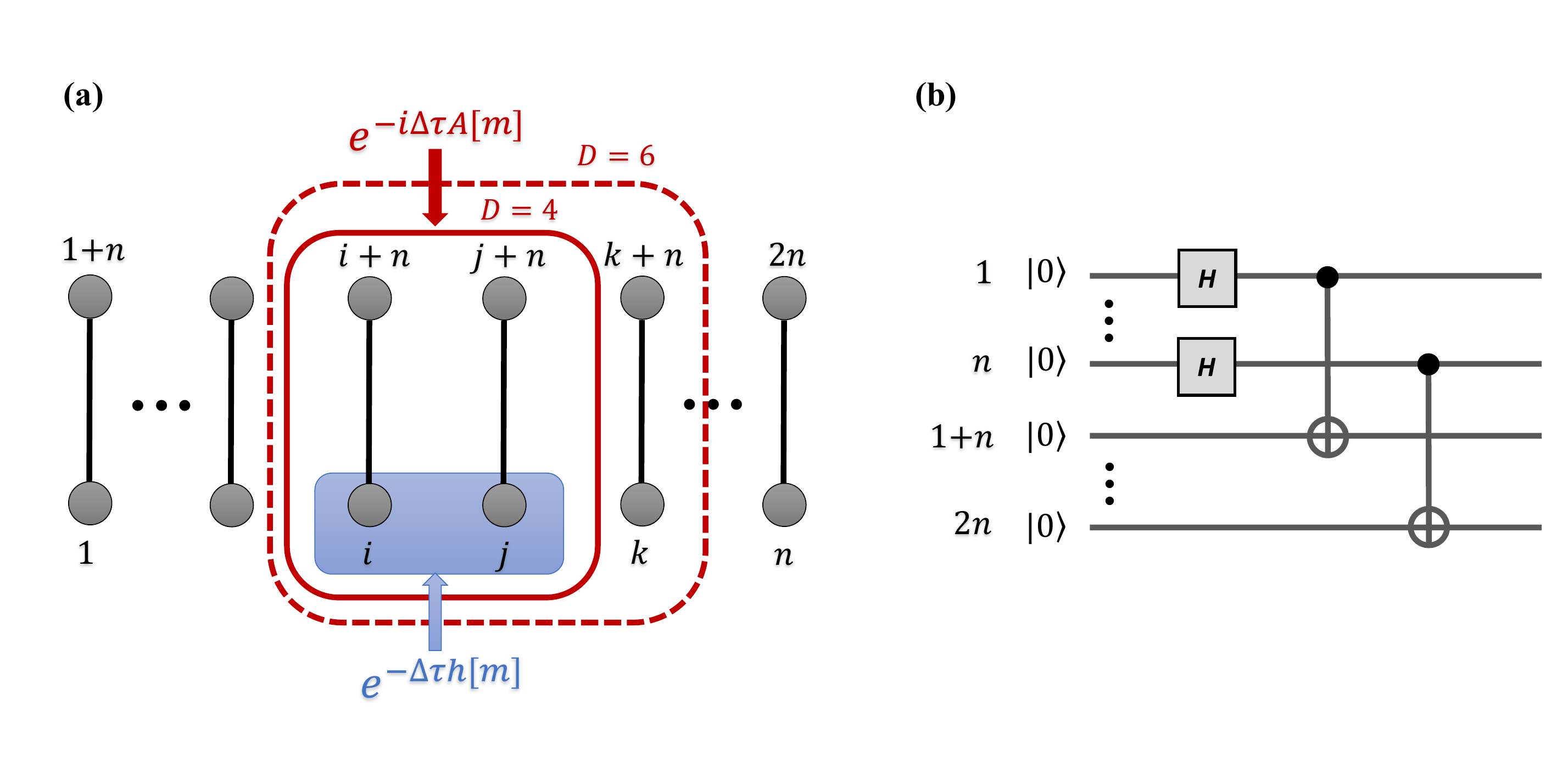}}
\caption{(a) Schematic illustration of the basic idea of QITE algorithm and the locality of unitary evolution in QITE  when apply to prepare the TFD state.  The qubits from $1$ to $2n$ form an one dimension chain. The bond between $i$ qubit and $i+n$ qubit indicate the  two qubit maximally entangled state $(|00 \rangle +|11 \rangle)/\sqrt{2}$ on $(i,i+n)$ qubits. And the one dimension chain represent the state $\ket{\text{TFD}(0)}$ on $1$ to $2n$ qubits before apply the quantum imaginary time transformation. The blue shadow frame indicates the $L$ ($L=2$) local  imaginary time transformation $e^{-\Delta\tau h[m]}$, which can be reproduced by unitary transformation $e^{-i\Delta\tau A[m]} $  acting on $D\geq L$ qubit. When the $L$ ($L=2$) local operator acting on $(i, j)$ qubits inside the blue shadow frame, the $D$ ($D=4$) qubit unitary operator  act on ${(i, j, i+n, j+n)}$ qubits indicated by the solid red box. When $D=6$ the unitary operator acting on ${(i,j,k, i+n,j+n,k+n)}$ qubits indicated by the dotted red box. (b) The quantum circuit to prepare the state $\ket{\phi_0}$ describe in the main text.}
\label{qite-locality}
\end{figure*}


{\textit{Step (S1)--}} The first major step (S1) of VQHD is to prepare quantum Gibbs state $\rho_{\beta}$, and to do so is 
known to be challenging~\cite{Aharonov2013}.
There have been some proposed quantum algorithms to prepare the thermal state. 
Some of the methods are based on quantum sampling~\cite{Temme2011QMS,Poulin2009Sampling,Yung2012QMA}, which require quantum phase estimation as a subroutine, therefore are not suitable for NISQ devices.

There are also some variational algorithms proposed to be implemented on NISQ devices~\cite{wu2019variational, Wang2020VQTS, Chowdhury2020VQTS, zoufal2020variational}. The methods used in Refs~\cite{wu2019variational, Wang2020VQTS, Chowdhury2020VQTS} 
are based on minimizing the free energy of the system at a certain temperature, which is challenging due to the nontrivial estimation of von Neumann entanglement entropy. Different authors use different approximation methods for the  free energy estimation for running their algorithms on NISQ devices. However, these methods also introduce other degrees of complexity.
Ref.~\cite{Chowdhury2020VQTS} approximately estimates the free energy with tools including quantum amplitude estimation~\cite{Brassard2000} and density matrix exponentiation~\cite{lloyd2014quantum,Hao2016Hamiltonian}. 
Ref.~\cite{Wang2020VQTS} estimates the approximate free energy with a truncated Taylor series. 

Two other proposals given in~\cite{wu2019variational, zoufal2020variational} both start from the thermofield double (TFD) state and need to evolve in imaginary time on a quantum device.  Ref.~\cite{wu2019variational} argues that it is hard to implement quantum imaginary time evolution and design a variational time evolution between the origin Hamiltonian and interacted Hamiltonian, which also needs to approximately estimate the nontrivial free energy with Renyi entropy estimation~\cite{Islam2015Measuring}. While Ref.~\cite{zoufal2020variational} straightforwardly employs a variational quantum imaginary time evolution (VarQITE)~\cite{McArdle2019QITE, Yuan2018Theory}, which makes the algorithm much simpler. All those variational methods share the drawback that it is hard to design the variational ansatz space to include the target point and the barren plateaus effect~\cite{McClean2018}.

The method starting from the TFD state and evolving in imaginary time for thermal state preparation is straightforward and simple, if we can implement the quantum imaginary time evolution (QITE) easily. Motta \textit{et al.}~\cite{motta2020determining} proposed an alternative QITE algorithm in addition to  the variational one~\cite{McArdle2019QITE}. The QITE can be applied to determine ground state energy and the thermal average. We adopt this QITE algorithm to prepare the thermal state, which can take advantage of the small correlation length of the system in small $\beta$. Consequently, our method for preparing the thermal state $\rho_{\beta}$  is to use the QITE algorithm given in \cite{motta2020determining} to evolve the TFD state as discussed in \cite{wu2019variational}, and then trace out the subsystem of the TFD state.

Consider a $2n$-qubit pure state $\ket{\text{TFD}(0)}$, which is a maximally entangled state between the first $n$ qubits (i.e. qubits $1,2,\ldots n$) and the other $n$ qubits (i.e. qubits $n+1,n+2,\ldots 2n$),
\begin{equation}
\ket{\text{TFD}(0)}= \sum_{i=1}^{2^n}\ket{i}\ket{i},
\end{equation}
where $\{\ket{i}\}$ are the computational basis of $n$ qubits.
We now define the state  

\begin{eqnarray}
\ket{\text{TFD}(\beta)}&=&\frac{e^{-\beta H/2}}{\sqrt{\tr(e^{-\beta H})}} \ket{\text{TFD}(0)}\nonumber\\
&=&\frac{1}{\sqrt{\tr(e^{-\beta H})}}\sum_{j=1}^{2^n} e^{-\beta h_j/2} \ket{\varphi_j}\ket{\varphi_j},
\end{eqnarray}
where $\{\ket{\varphi_j}\}$ are the orthonormal eigenvectors of $H$, and $\{h_j\}$ are the corresponding eigenvalues.
Tracing out the qubits $n+1,n+2,\ldots, 2n$ from $\ket{\text{TFD}(\beta)}$ gives
\begin{equation}
\rho_{\beta}=\tr_{n+1,n+2,\ldots, 2n}\ket{\text{TFD}(\beta)}\bra{\text{TFD}(\beta)}.
\end{equation}

To be able to implement the quantum imaginary time evolution $e^{-\beta H/2}$ on a quantum computer, we use a QITE algorithm as given in Ref.~\cite{motta2020determining}. 
After Trotter decomposition of the corresponding imaginary time evolution in $N=\frac{\beta/2}{\Delta\tau}$ steps, the basic idea of QITE is map the $L$-local non-unitary transformation to an approximate $D$ local unitary transformation in each step,
\begin{eqnarray}
|{\psi}' \rangle = \frac{1}{\sqrt{c}} e^{-\Delta\tau h[m]} |\psi \rangle \approx e^{-i\Delta\tau A[m]} |\psi \rangle, 
\label{basicQite}
\end{eqnarray}
where $c = \langle \psi|e^{-2\Delta\tau h[m]} |\psi \rangle$ is the normalization factor. Each $h[m]$ acts nontrivially on $L$ qubits. $A[m]$ is Hermitian and act on $D$ qubits. $A[m]$ can be expanded in terms of Pauli basis on $D$ qubits,
\begin{eqnarray}
A[m] = \sum_{i_1 i_2 ... i_D} a[m]_{i_1 i_2 ... i_D} \sigma_{i_1} \sigma_{i_1} ... \sigma_{i_D} = \sum_{I} a[m]_{I} \sigma_{I},
\end{eqnarray}  
where $a[m]_{I}$ is the coefficient of combine Pauli operator $\sigma_{I}$ and the index $I$ is a combination of qubit indexes$\{i_1,i_2, ... ,i_D\}$. To find coefficients $a[m]_{I}$ and determine the concrete form of $A[m]$, we minimize the function 
\begin{eqnarray}
\parallel |{\psi}' \rangle -(1-i\Delta\tau A[m]) |{\psi} \rangle \parallel, 
\end{eqnarray}
which is consistent with our goals as discribed in Eq.~\eqref{basicQite}. It can be easily derived that the solution of the minimization is subject to the linear equation,
\begin{eqnarray}
\label{linear-eq}
(\boldsymbol{S} + \boldsymbol{S}^T) \boldsymbol{a}[m] = -\boldsymbol{b},
\end{eqnarray}
where the matrix $\boldsymbol{S}$ and vector $\boldsymbol{b}$ can be obtained by $D$ local measurements on the $|{\psi} \rangle$,
\begin{eqnarray}
\label{eq:Sb-matrix}
S_{IJ} &=& \langle \psi| \sigma_{I}^{\dagger} \sigma_{J}  |\psi \rangle , \nonumber \\
b_{I}  &=& -  2\text{Im}\left[\frac{1}{\sqrt{c}}\langle \psi| \sigma_{I}^{\dagger} h[m] | \psi \rangle \right],
\end{eqnarray}
where Im[] means the imaginary part of the variable inside.
After solve the  linear equation in the classical computer, we get the $\boldsymbol{a}[m]$ and construct a quantum circuit to implement the unitary transformation $e^{-i\Delta\tau A[m]} |\psi \rangle$ on NISQ  quantum devices at each step. 

One of the most important parameters of the QITE algorithm is  $D$, which is the number of qubits that the local unitary transformation acts on. Given $L$ local Hamiltonian, the QITE algorithm can captures the correlation of the original Hamiltonian only if  $D\geq L$.  However, our goal is to prepare the thermofield double state $\ket{\text{TFD}(\beta)}$ start from $\ket{\text{TFD}(0)}$, which is a maximally entangled state between the first $n$ qubits and the other. When the unitary local operator act on $i$ qubit, it must include the $i+n$ qubit and $D\geq 2L$ . In this paper, we consider the 2 local Hamiltonian and $D \geq 4$. The locality of the $D$ qubit unitary operator  in QITE algorithm is shown in Fig. \ref{qite-locality} (a) with $L=2$, $D=4$ and $D=6$.

Notes that the thermofield double state  $\ket{\text{TFD}(0)}$ is not normalized. We define a variable
\begin{eqnarray}
\ket{\phi_0}  = \frac{1}{\sqrt{2^n}} \ket{\text{TFD}(0)} = \frac{1}{\sqrt{2^n}} \sum_{i=1}^{2^n}\ket{i}\ket{i}.
\label{eq:inital-state}
\end{eqnarray}
The $\ket{\phi_0} $ is normalized and can be easily prepared with a quantum circuit. Fig.~\ref{qite-locality} (b) shows the quantum circuit to prepare the state $\ket{\phi_0}$: start with the initial state $|0\rangle^{\otimes 2n}$ and apply Hadamard gates on qubits from $1$ to $n$; then apply the CNOT gates on $(i, i+n)$ qubits, where $i$ is the control qubit and $i+n$ is the target qubit and $i$ run over $1$ to $n$.  After preparing the initial state $\ket{\phi_0} $ by a quantum circuit,  we use the QITE algorithm to obtain 
\begin{eqnarray}
\ket{\text{TFD}(\beta)} =  \sqrt{\frac{2^n}{\tr(e^{-\beta H})}} e^{-\beta H/2} \ket{\phi_0}.
\label{ite}
\end{eqnarray}

\begin{figure*}[tb]
\centerline{\includegraphics[width=0.85\textwidth]{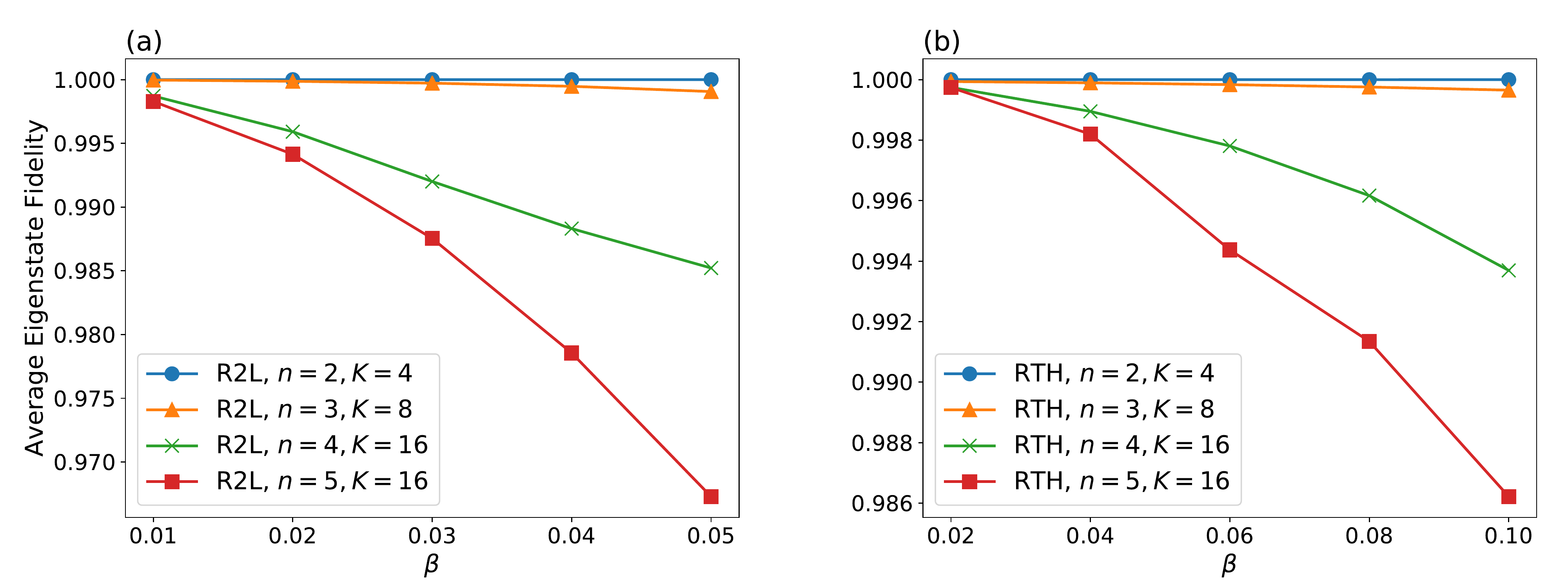}}
\caption{The average fidelity between the predicted K eigenstates from VQHD and the exact lowest K eigenstates of H for \textbf{(a)} random 2-local Hamiltonian H  and \textbf{(b)} random transverse field Heisenberg Hamiltonian. The performance of VQHD declines as $n$ and $\beta$ increase. But for small $\beta$, VQHD always performs well. }
\label{VQSE-AF}
\end{figure*}

{\textit{Step (S2)--}}  The second major step (S2) of VQHD is to diagonalize $\rho_{\beta}$ for obtaining the eigenvalues and eigenvectors of $\rho_{\beta}$.
In general, for any quantum state $\rho$, quantum principle component analysis~\cite{lloyd2014quantum} can be used to diagonalize $\rho$ with an exponential speedup compared to classical computers. However, this method needs fault-tolerance hence it cannot be implemented on NISQ devices. Variational Quantum State Diagonalization (VQSD)~\cite{larose2019variational} and Variational Quantum State Eigensolver (VQSE)~\cite{cerezo2020variational} are alternative algorithms to extract the eigenvalues and eigenstates of $\rho$, and they can be used on near-term devices. Compared with VQSD, VQSE needs fewer qubits. Here we then use VQSE for diagonalizing $\rho_{\beta}$.

After preparing the thermal state $\rho_{\beta}$ with QITE, we train a parameterized quantum circuit $V(\boldsymbol{\theta})$ to partially diagonalize it with VQSE. Denote the circuit output state as $\rho_{f}$,  $\rho_{f} = V(\boldsymbol{\theta}) \rho_{\beta} V^{\dagger}(\boldsymbol{\theta})$. Suppose we want to obtain the lowest $K$ eigenstates of $H$, the cost Hamiltonian is therefore
\begin{equation}\label{hcost}
H_{cost} \equiv \mathbb{1}-\sum_{i=1}^{K} q_{i}\left|i\right\rangle\left\langle i\right|,
\end{equation}
where $\{\ket{i}\}$ are the computational basis and $q_{i} > q_{i+1}$ for $1 \leq i \leq K-1$. We measure the expectation value of $\rho_f$ on the cost Hamiltonian to estimate the cost function

\begin{equation}\label{loss-vqse}
C(\boldsymbol{\theta}) =\operatorname{Tr}\left[H_{cost}\rho_{f}\right],
\end{equation}then optimize the parameters $\boldsymbol{\theta}$ to minimize it.  We use gradient descent methods to trained the parameterized quantum circuits. Although the gradient will vanish exponentially as $n$ increase~\cite{McClean2018}, there are several methods to suppress the phenomenon, such as selecting special initial parameters~\cite{grant2019initialization}, replacing the global cost function by a local cost function~\cite{cerezo2020cost}, training the circuit with an adaptive Hamiltonian~\cite{cerezo2020variational}. 

$\rho_{\beta}$ and $H$ share the same eigenstates. $\rho_{\beta}$ of a non-degenerate Hamiltonian is also non-degenerate, reaching the minimum value of $C(\boldsymbol{\theta})$ indicates an exact partial diagonalization of the  $\rho_{\beta}$. Denote the eigenstates of $H$ as $\{|\psi_{k}\rangle\}$, where $k$ is the level index. $V(\boldsymbol{\theta})|\psi_{k}\rangle\ = |k\rangle$ for exact partial diagonalization with $k \leq K$, we can run the inverse of $V(\boldsymbol{\theta})$ to prepare the lowest $K$ eigenstates of $H$, measurements of $H$ give the corresponding eigenvalues,
\begin{equation}
\langle \psi_{k} |H| \psi_{k} \rangle = \lambda_{k}.
\end{equation}


\subsection{Simulation results}
We implement the VQHD algorithm to demonstrate its feasibility. 
We apply our the algorithm to diagonalize two one-dimensional Hamiltonian, 
the random 2-local Hamiltonian and the random transverse field Heisenberg Hamiltonian with periodic boundary condition. 

The one-dimensional random  2-local Hamiltonian is defined as
\begin{eqnarray}
H_{R2L} & = &  \sum_{\langle ij \rangle}  \boldsymbol{h}[i] \cdot  \boldsymbol{\sigma} [i,j]  \nonumber  \\ 
              & = &  \sum_{\langle ij \rangle} \sum_{I}^{16} {h}_{I}[i] \sigma_{I} [i,j]   \nonumber  \\ 
              & = &  \sum_{\langle ij \rangle} \sum_{\alpha=1}^{4} \sum_{\beta=1}^{4} {h}_{\alpha\beta}[i] \sigma_{\alpha} [i] \sigma_{\beta}[j],
\end{eqnarray}
where $\langle ij \rangle$ denote the summation over the nearest-neighbor (NN) lattice site. $\alpha, \beta$ denote the Pauli operator index and $\sigma_{\alpha}[i] $ is one of the Pauli operators $(\sigma_{0}  = \text{I}, \sigma_{1}  = \text{X}, \sigma_{2}  = \text{Y}, \sigma_{3}  = \text{Z})$ act on site $i$.  The coefficient ${h}_{\alpha\beta}[i]$  is sampling from a uniform distribution over the interval $[0, 1)$ and subject to the condition $\sum_{i\alpha\beta} {h}_{\alpha\beta}[i] = 16n$, where $n$ is the number of lattice sites.

The one-dimensional Heisenberg Hamiltonian with random transverse field is defined as
\begin{eqnarray}
H_{RTH} =  \sum_{\langle ij \rangle} \boldsymbol{S}_{i} \cdot \boldsymbol{S}_{j} + \sum_{i}^{n} h_i Z_{i},
\end{eqnarray}
where $\langle ij \rangle$ denotes the summation over the nearest-neighbor (NN) lattice site. $h_i$ is the coefficient of the transverse field term on each site, which is sampling from a uniform distribution over the interval $[0, 1)$ and subject to the condition $\sum_{i} h_i = n$, where $n$ is the number of lattice sites.


The results for two Hamiltonian are shown in Fig. \ref{VQSE-AF}. We diagonalize $H$ for $n = 2, 3, 4$, partially diagonalize $H$ for $n = 5$. A small $K$ that satisfies $K<<2^{n}$ makes the training much easier. We define the average fidelity between predicted $K$ eigenstates $\{| \psi_k \rangle \}$ from VQHD and the exact lowest $K$ eigenstate $\{| \psi_k^{e} \rangle \}$  of $H$ as $\frac{1}{K}\sum_{k=1}^{K}  | \langle \psi_k | \psi_k^e \rangle |^2$. We can see that the VQHD performs well when $\beta$ up to 0.05 for random 2-local hamiltonian and 0.1 for random transverse field Heisenberg hamiltonian. 

\section{Algorithm analysis}

In this section, we analyze the performance of our VQHD algorithm and compare to other related methods.
We will focus on the state of thermal state preparation.

\subsection{The performance of thermal state preparation}

\begin{figure*}[tb]
\centerline{\includegraphics[width=1.0\textwidth]{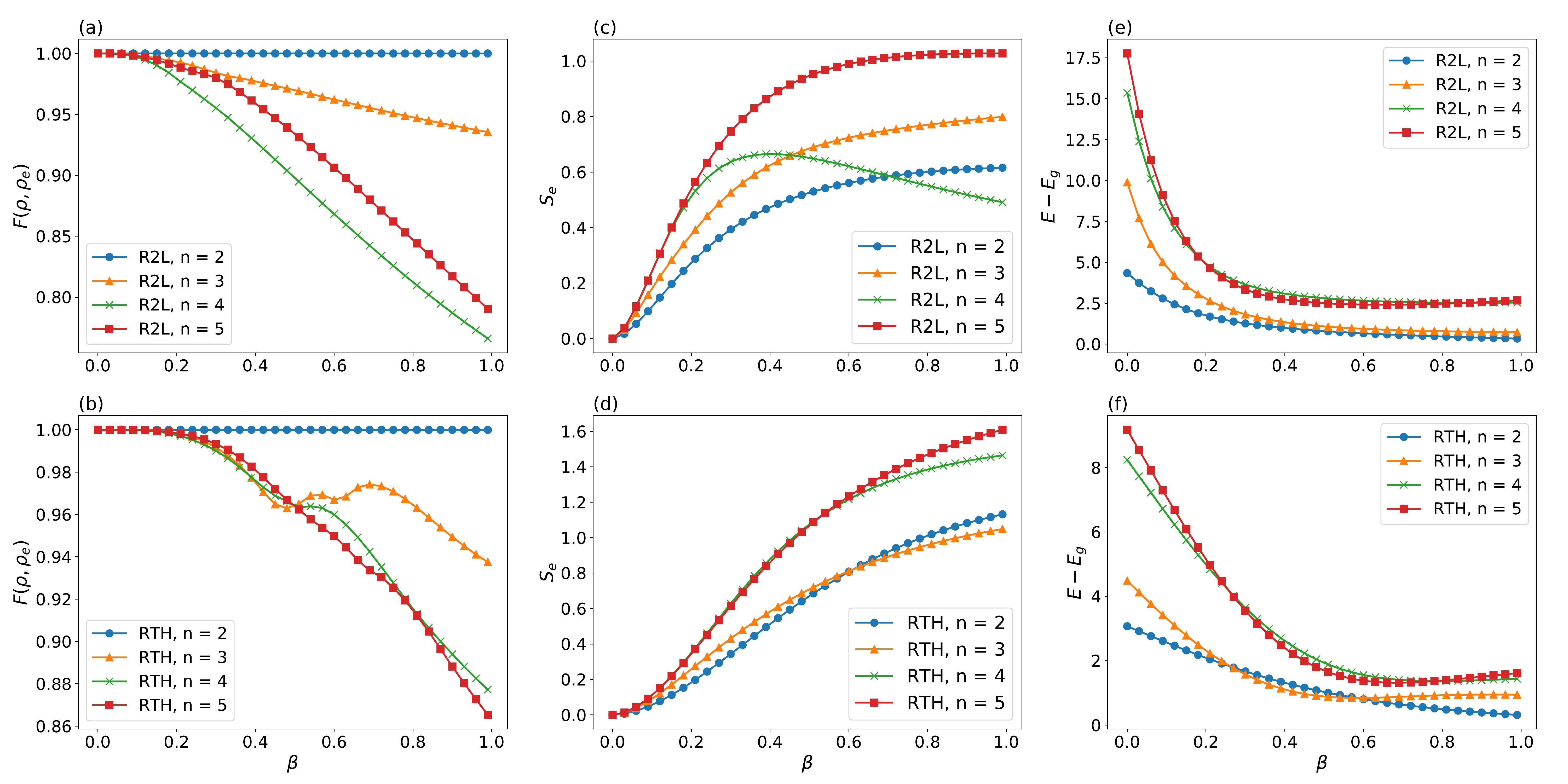}}
\caption{The numerical results of preparation of $\rho_{\beta}$ for random 2-local Hamiltonian and the random transverse field Heisenberg Hamiltonian with $D=4$. The fidelity between the $\rho$ produced by QITE and the exact $\rho_{0}$ as a function of $\beta$ for \textbf{(a)} random 2-local Hamiltonian and \textbf{(b)} random transverse field Heisenberg Hamiltonian. The von Neumann entropy for the exact state $|\psi_{CD} \rangle$ calculated by Eq.~\ref{ite} for \textbf{(c)} random 2-local Hamiltonian and \textbf{(d)} random transverse field Heisenberg Hamiltonian. The $C$ subsystem include the sites $(1,2, ..., n/2, 1+n, 2+n,  ..., n/2 + n)$. The relative energy $E-E_{g}$ between the energy calculated by QITE and the ground state energy $E_g$ as a function of $\beta$ respectively for \textbf{(e)} random 2-local Hamiltonian and \textbf{(f)} random transverse field Heisenberg Hamiltonian.}
\label{QITE-FIG}
\end{figure*}

In the part we discuss the  performance of preparing thermal state. The  accuracy of the thermal state can affects the performance of the VQHD algorithm.  As discussed in Section~IIB, the minimum $D$ is $4$ for preparing $\rho_{\beta}$. In all the experiments we set $D=4$. The torrter step interval $\Delta\tau$ is $0.005$. We use the expectation values give by Eq.~\ref{eq:Sb-matrix} to construct the matrix $S$ and vector $b$. 

The results of the random 2-local Hamiltonian and the random transverse field Heisenberg Hamiltonian are summarized in Fig.~\ref{QITE-FIG}.  To characterize how well the QITE can prepare the  target state $\rho_{\beta}$, Fig.~\ref{QITE-FIG} (a) and (b) shows $F(\rho, \rho_{e})=  \left( \tr \sqrt{\sqrt{\rho}\rho_e\sqrt{\rho}}\right)^2,$ 
the fidelity between the $\rho$ produced by QITE and the exact $\rho_{e}$ as a function of $\beta$ respectively for (a) random 2-local Hamiltonian and (b) random transverse field Heisenberg Hamiltonian. When the Hamiltonian defined on $n=2$ site, the whole $\ket{\text{TFD}(\beta)}$ defined on $2n=4$ qubits. If we set $D=4$, the unitary operator act on the whole system and can capture the correlation of system. The fidelity $F(\rho, \rho_{e})$ of $n=2$ is approximated to 1 in all $\beta$ as shown in the blue point line in Fig.~\ref{QITE-FIG} (a) and (b). When $n>2$ the fidelity $F(\rho, \rho_{e})$ decrease as $\beta$ increase. However, the fidelity $F(\rho, \rho_{e})$ still approximate 1 for $\beta < 0.1$ and state produced by QITE preform well in small $\beta$. This could be understood as follows, for small $\beta$ the many body state $\rho_{\beta}$ is near the initial $\rho_{0}$ and exhibit small correlation. Fig. \ref{QITE-FIG} (c) and (d) show the  von Neumann entropy $S_{e} = \tr{\rho_{C}\log \rho_{C}}$ for the exact state $|\psi_{CD} \rangle$ calculated by the Eq.~\ref{ite} respectively for (c) random 2-local Hamiltonian and (d) random transverse field Heisenberg Hamiltonian, where subsystem $C$  include the sites $(1,2, ..., n/2, 1+n, 2+n,  ..., n/2 + n)$. The entropy $S_{e}=0$ at $\beta=0$ and the entropy increase as $\beta$ increase. The unitary act only $D=4$ qubits can approximate the small correlation of the original Hamiltonian for small $\beta$, but can not capture the correlation for large $\beta$. 

Note that when applying for the QITE to solve the ground state energy problem. We need large $\beta$ and large $D$ for big $n$.  Fig. \ref{QITE-FIG} (e) and (f) shown the relative energy $E-E_{g}$ between the energy calculated by QITE and the ground state energy $E_g$ respectively for (e) random 2-local Hamiltonian and (f) random transverse field Heisenberg Hamiltonian.  As $\beta$ increase to $1$, the relative energy are convergent for all $n$. But only the energy of $n=2$ converges to ground state energy. The relative energy $E-E_g$ is large for large $n$. Compare to the ground state problem, we need a small $D$ and small $\beta$ to prepare the $\rho$ when apply for the QITE algorithm. As space and time cost of QITE are proportional to exponentials in $D$. The number of measurements is also bounded by $e^{D}$. Hence, we reduce the time and space cost as well as the number of measurements.

\begin{figure*}[tb]
\centerline{\includegraphics[width=1.0\textwidth]{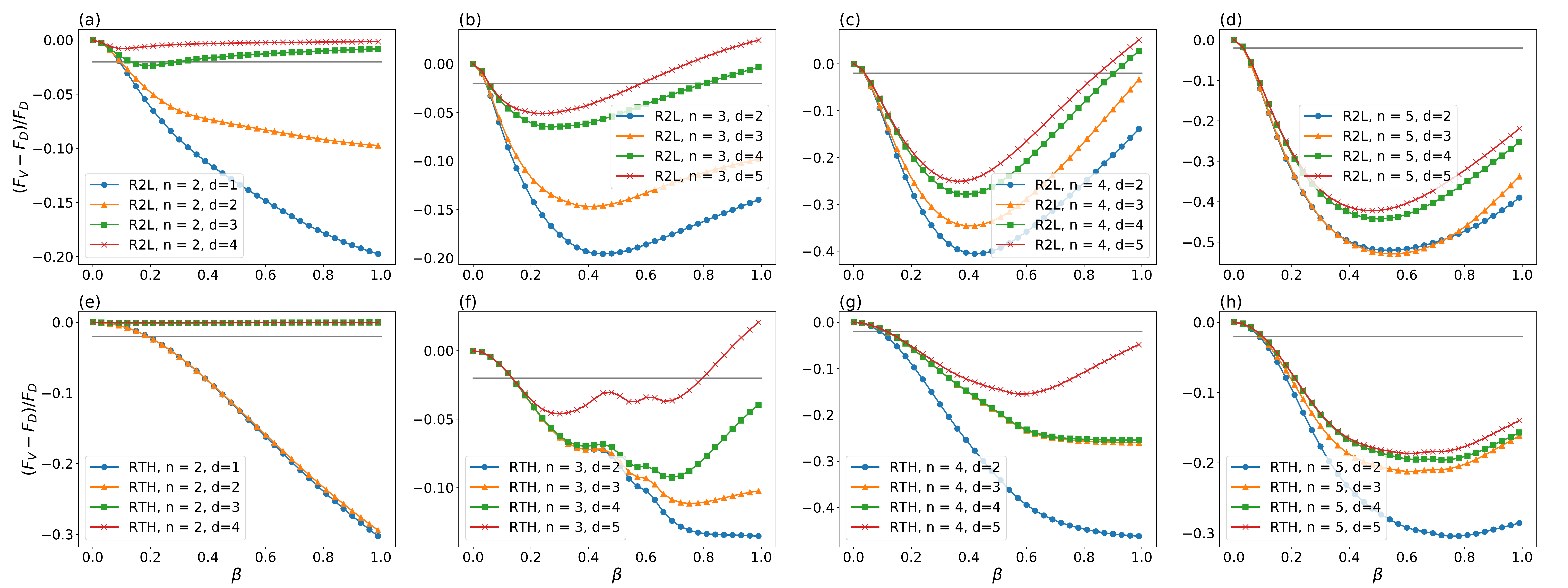}}
\caption{The comparison of QITE and VarQITE for preparing thermal state. The relative fidelity $(F_{V} - F_{D})/F_{D}$ between the fidelity $F_{V}$ calculated by VarQITE  and the fidelity $F_{D}$ calculated by QITE for random $2$-local Hamiltonian (R2L) with 2 (a), 3 (b), 4 (c), 5 (d) spins.  $(e-f)$ are the results of random transverse field Heisenberg Hamiltonian (RTH). $F_{V}$ is the fidelity between the $\rho$ produced by VarQITE and the exact $\rho_0$. $F_{D}$ is the fidelity between the $\rho$ produced by QITE and the exact $\rho_0$. In each $n$ spin model, we run several circuit depth $d$ for VarQITE to explore the capacity of the circuit ansatz. The grey horizontal lines represent the level -2\%. The relative fidelity $(F_{V} - F_{D})/F_{D} < 0$ for most of case in small $\beta$ region, say $\beta \in [0, 1]$, mean that the QITE outperform the VarQITE for preparing the thermal state in small $\beta$.  }
\label{fig:compare-QITE-VQITE}
\end{figure*}

\subsection{Comparison of QITE and VarQITE  for thermal state preparation}

The main reason we use QITE instead of VarQITE to prepare the thermal state $\rho_{\beta}$ is that the QITE algorithm takes advantage of the small correlation length of the system for small $\beta$. As shown in Fig.~\ref{QITE-FIG} (c) and (d), the correlation length of $\rho_{\beta}$ for random $2$-local Hamiltonian and random transverse field Heisenberg Hamiltonian is small for small $\beta$. The QITE algorithm can restrict the local unitary transformations on the relatively small size (i.e. small $D$) hence saves the quantum resource and is easy to run on NISQ quantum computers. However, the VarQITE~\cite{McArdle2019QITE} algorithm always need to construct a global quantum circuit ansatz, which is a subspace of the full Hilbert space. It is a challenge to design an efficient ansatz with small parameters and low depth circuits. In this section, we compare the performance and the corresponding quantum resource requirements of  QITE and VarQITE for preparing $\rho_{\beta}$. To be specific we estimate the number of Pauli measurements needed for QITE and VarQITE in the small $\beta$ regime.

The number of Pauli measurements in each iteration of QITE is $N_m \times 4^{D}$, where $N_m$ is the number of Pauli terms in the Hamiltonian, and $D$ depends on the correlation length. While the number of  measurements  of 
VarQITE is $N_{p}^{2} + N_{p}  \times N_{m} $~\cite{McArdle2019QITE}, where $N_{p}$ is the number of parameters.
For geometrically local spin lattice models, $N_{m} =\mathcal{O}(n)$, where $n$ is the number of spins.  $N_{p}$ depends on the quantum circuit ansatz. When using hardware efficient ansatz, $N_{p} = \mathcal{O}(nd) $, where $d$ is the circuit depth. Then the number of measurements become $\mathcal{O} (n \times 4^{D})$ for QITE and $\mathcal{O}  \left(n^2 \times(d^2 + d)\right)$ for VarQITE. The number of measurements depends on $\mathcal{O}(n^2)$ in VarQITE, which is worse than $\mathcal{O}(n)$ dependence in QITE.

It is hard to generally compare $D$ in QITE and the circuit depth $d$ in VarQITE. Then we come to the particular cases of random 2-local Hamiltonian and random transverse field Heisenberg model. In all the simulations of QITE, we set $D=4$.  In the simulations of VarQITE, we construct the whole circuit ansatz with the circuit described in Ref.~\cite{zeng2020simulating} followed by the circuit described in Fig.~\ref{qite-locality} (b) but replace the Hadamard with $R_{z}(\theta_1)R_{x}(\theta_2)R_{z}(\theta_3)$ in the first $n$ qubits. As discussed in Ref.~\cite{zeng2020simulating}, the circuit has suitable expressibility and entangling capability for the one-dimensional spin models. The tail circuit layer added in the end is for preparation the maximally entangled state $\ket{\phi_0}$ (Eq.~\ref{eq:inital-state}) as the imaginary time evolution should begin with the maximally entangled state. The whole circuit ansatz for VarQITE has $N_{p} = 4\times(2n-1)d + 3n$ tunable parameters. When we set the first $4\times(2n-1)d$ parameters as zeros and $\theta_1 = 3\pi/2, \theta_2 = \pi/2 , \theta_3 = 5\pi/2$ for the first $n$ qubits in the tail layer, the output state is $\ket{\phi_0}$. Then we can use VarQITE to prepare the thermal state $\rho_{\beta}$.

\begin{table}[th]
\centering
\addtolength{\tabcolsep}{9pt}
\caption{ The total number of measurements of QITE and VarQITE for the random 2-local Hamiltonian (R2L) and random transverse field Heisenberg Hamiltonian (RTH).  $n$ is the number of spins. $N_m$ is the number of Pauli terms in the Hamiltonian. $d$ is the minimal circuit depth for VarQITE to have the same performance as QITE in $\beta = 1$ within $2\%$. For 5 spin random 2-local Hamiltonian,  and 4, 5 spin random transverse field Heisenberg Hamiltonian, we set $d=5$. In these cases the VarQITE still can not outperform QITE, however, have a larger number of Pauli measurements than QITE.}
\label{tab:number-measurement}
\begin{tabular}{ccccccc}
\toprule
 $n$   &   2 & 3   & 4 &  5  \\ \midrule
\textbf{R2L} \\ 
$N_{m}$    & 16  & 48 & 64 & 80 \\
$d$             &  3 &   4  & 4 & 5  \\
VarQITE  & 2436  & 12193  & 23312  & 53625   \\ 
QITE      & 4096  & 12288  & 16384 & 20480  \\ \midrule
\textbf{RTH} \\ 
$N_{m}$    & 5  & 12 & 16 & 20 \\
$d$             &  3 &   5  & 5 & 5  \\
VarQITE  & 1974  & 13189  & 25536 & 41925   \\ 
QITE    & 1280  & 3072  & 4096 & 5120   \\ \bottomrule
\end{tabular}
\end{table}

We prepare the thermal state of one dimensional random $2$-local Hamiltonian and random transverse field Heisenberg model with $2, 3, 4, 5$ spins. In each $n$ spin model, we run VarQITE simulations with several different circuit depth $d$ and compare with QITE with $D=4$. The $y$-axis in Fig.~\ref{fig:compare-QITE-VQITE}  represent the relative fidelity $(F_{V} - F_{D})/F_{D}$ between the fidelity $F_{V}$ calculated by VarQITE  and the fidelity $F_{D}$ calculated by QITE. From Fig.~\ref{fig:compare-QITE-VQITE}, we can see that the QITE outperforms the VarQITE for preparing the thermal state in small $\beta$. As the circuit depth $d$ increases, the performance of VarQITE increases. If the circuit ansatz has enough capacity, the VarQITE performs as well as QITE, as shown in Fig.~\ref{fig:compare-QITE-VQITE} (a) for the $2$ spin random $2$-local Hamiltonian with $4$ circuit depth and in Fig.~\ref{fig:compare-QITE-VQITE} (e) for the 2 spin random transverse field Heisenberg Hamiltonian with $3$ circuit depth. We then compare the quantum resources needed for QITE and VarQITE. We increase the circuit depth in VarQITE and find the minimal $d$ which satisfies the condition $(F_V - F_D)/F_D > -2\%$ at $\beta = 1$ and then calculate the number of Pauli measurements. The results are summarized in Tab.~\ref{tab:number-measurement}. The QITE  algorithm can save the number of Pauli  measurements comparing to the VarQITE for the random transverse field Heisenberg model in small $\beta$. The same conclusion can holds in the case of random $2$-local Hamiltonian, even though the number of Pauli measurements in QITE are a little bit larger than the value in VarQITE for the case of $n=2$ and $n=3$.

\section{Discussion}

In this work, we proposed a variational algorithm for diagonalizing many-body Hamiltonians. Our VQHD method
explores important physical properties of local Hamiltonians, including temperature, locality and correlation.

The key idea is that the thermal state $\rho_{\beta}=e^{-\beta H}/\tr (e^{-\beta H})$ encodes the information of eigenvalues and eigenstates
of $H$, for any $\beta$ in principle. Due to the exponential function, larger $\beta$ will suppress high energy states, hence to retrieve information of low-lying states of $H$. Diagonalizing $\rho_{\beta}$ with a variational algorithm on a quantum computer
will then result in either full spectrum or low-lying eigenstates of $H$, depending on $\beta$.

$\rho_{\beta}$ can be obtained from the thermofield double state $\ket{\text{TFD}(\beta)}=e^{-\beta H/2}\ket{\text{TFD}(0)}$, by replacing the imaginary time evolution $e^{-\beta H/2}$ with a unitary transformation implemented on a quantum computer. For small $\beta$, the many-body state $\ket{\text{TFD}(\beta)}$ has a small correlation length. With Trotterization of the imaginary time evolution $e^{-\beta H/2}$, each step can hence be replaced by a unitary transformation on only a small number of sites. 

We also analyze the performance of QITE for preparing the thermal state. Our results suggest that the QITE algorithm can take advantage of the small correlation length of the system in small $\beta$ and hence save the quantum resources. To solve the ground state energy, we should go to the large $\beta$ region. In principle, starting from an initial state that has some overlap with the ground state,  evolve imaginary time to infinity large $\beta$ can converges to the ground state. The QITE algorithm needs a large $D$  to capture the large correlation length in the ground state of the system. It is worth comparing the performance and quantum resources requirementsof QITE and VarQITE for the ground-state problem. Another significant difference between the thermal state preparation and the ground state problem is the locality of the unitary operators in QITE.  We say that the TFD state has a smaller correlation length for small $\beta$, which is just defined on the first n qubits. Note that, the initial state $\ket{\text{TFD}(0)}$  is a maximally entangle state between the first $n$ qubits and final $n$ qubits. The unitary transformations must capture the correlations between the $i$ and $i+n$ qubits, hence always need to include the pair $(i, i+n)$  in the $D$ qubits unitary transformations.

Our VQHD algorithm introduces a hyperparameter $\beta$, which can balance accuracy and complexity between the QITE for preparation of thermal state and the VQSE for diagonalization of thermal state depending on different problems. The time complexity of the QITE and the depth of the circuit construct by the QITE algorithm depend on $N_{T} \times \mathcal{O}(e^{D}) $, where $N_{T}$ is the number of Trotter steps. Since the small correlation of the thermal state in the small $\beta$ region, we can use a small constant  $D$ in QITE ( $D=4$ in our cases). When  fixing the interval of the Trotter step $\Delta\tau$,  the time complexity and the final whole circuit depth of QITE only depend on $\beta$. Hence, the QITE  favors small $\beta$, with small time complexity, small circuit depth, small Trotter error. However, the VQSE does not favor small $\beta$, since the loss function in Eq.~\ref{loss-vqse} is flat when $\beta$ is very small. The optimization procedure is hard, which impacts the performance of the VQSE and the time complexity of the optimization. This problem can be mitigated by using a better classical optimizer and using adaptive cost hamiltonian \cite{cerezo2020variational} instead of the fixed cost hamiltonian in Eq.~\ref{hcost}. When the $\beta$ increase, the loss function becomes sharp and is easy to optimize. In the VQHD algorithm, we should choose the relatively large value in the small $\beta$ region which can ensure that the QITE can prepare  fidelity thermality with constant $D$. Finally, it is worth to developing of quantum algorithm which effectively diagonalizes thermal state at high temperature or quantum algorithms to prepare thermal state at low temperature.


\section*{Acknowledgement}
PX acknowledges the support by the Key-Area Research and Development Program of Guangdong Province (No.2019B121204008).

\bibliographystyle{apsrev4-1}
\balance
\footnotesize
\bibliography{VQHD}

\end{document}